\documentclass[prb,twocolumn,a4paper,showpacs,superscriptaddress]{revtex4}

\usepackage{graphicx}
\usepackage{amsmath}
\usepackage{amssymb}
\usepackage{amsfonts}
\usepackage{color}
\usepackage[usenames,dvipsnames]{xcolor}

\begin{document}

\title{Magnetic ordering in GdNi$_{\mathrm{2}}$B$_{\mathrm{2}}$C revisited by resonant x-ray scattering: evidence for the double-$q$ model}
\date{\today}
\author{P. S. Normile}
\thanks{peter.normile@uclm.es}
\affiliation{Instituto Regional de Investigaci{\'{o}}n Cient{\'{i}}fica Aplicada (IRICA) and Departamento de F{\'{i}}sica Aplicada, Universidad de
Castilla-La Mancha, 13071 Ciudad Real, Spain} \affiliation{Department of Physics, University of Liverpool, Liverpool L69 3BX, United Kingdom and XMaS, European Synchrotron Radiation Facility, 38043 Grenoble, France}
\author{M. Rotter}
\affiliation{Max-Planck Institute for Chemical Physics of Solids, 01187 Dresden, Germany}
\author{C. Detlefs}
\affiliation{European Synchrotron Radiation Facility, BP 220, 38043 Grenoble Cedex 9, France}
\author{J. Jensen}
\affiliation{Niels Bohr Institute, Universitetsparken 5, DK-2100 Copenhagen, Denmark}
\author{P. C. Canfield}
\affiliation{Ames Laboratory, US DOE, and Department of Physics and Astronomy, Iowa State University, Ames, Iowa 50011, USA}
\author{J. A. Blanco}
\affiliation{Departamento de F{\'{i}}sica, Universidad de Oviedo, E-33007 Oviedo, Spain}

\begin{abstract}
Recent theoretical efforts aimed at understanding the nature of antiferromagnetic ordering in GdNi$_{\mathrm{2}}$B$_{\mathrm{2}}$C predicted double-$q$ ordering. Here we employ resonant elastic x-ray scattering to test this theory against the formerly proposed, single-$q$ ordering scenario. Our study reveals a satellite reflection associated with a {\em mixed}-order component propagation wave vector, viz., ($q_a$,2$q_b$,0) with $q_b$ = $q_a$ $\approx$ 0.55 reciprocal lattice units, the presence of which is incompatible with single-$q$ ordering but is expected from the double-$q$ model. A (3$q_a$,0,0) wave vector (i.e., third-order) satellite is also observed, again in line with the double-$q$ model.  The temperature dependencies of these along with that of a first-order satellite are compared with calculations based on the double-$q$ model and reasonable qualitative agreement is found. By examining the azimuthal dependence of first-order satellite scattering, we show the magnetic order to be, as predicted, elliptically polarized at base temperature and find the temperature dependence of the ``out of $a$-$b$ plane'' moment component to be in fairly good agreement with calculation. Our results provide qualitative support for the double-$q$ model and thus in turn corroborate the explanation for the ``magnetoelastic paradox'' offered by this model.

\end{abstract}


\maketitle

\section{Introduction}
\begin{figure}[!hbt]
\includegraphics[scale=.2,angle=0]{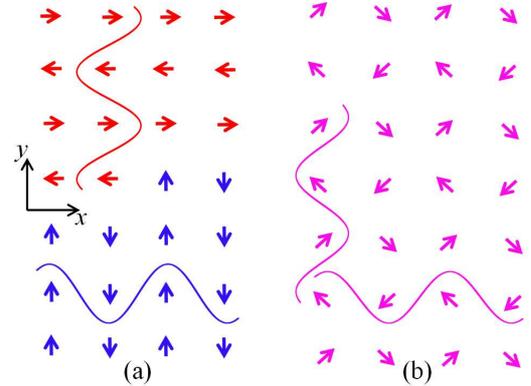}
\caption{(a) Single-$q$ and (b) double-$q$ representations of AFM order associated with a commensurate (principal) propagation wave vector, of magnitude $q$ = 0.5 reciprocal lattice units. The arrows represent ordered magnetic moments positioned in a square array. The lattice parameter is the separation between two adjacent moments.} \label{schematics_1q_2q}
\end{figure}

A well-known complexity in the determination of antiferromagnetic (AFM) structure arises in systems with a high symmetry (e.g., cubic or tetragonal) crystal lattice: the issue of whether the magnetic correlations in each AFM domain are associated with a single magnetic propagation wave vector axis or with multiple axes;\cite{Rossat_chapter_expt_meths} in other words, whether the domains are single-$q$ or multi-$q$. An illustrative example is depicted in Fig. \ref{schematics_1q_2q}. Panel (a) shows two-dimensional representations of a pair of orthogonal single-$q$ domains (blue and red arrows). The coherent sum of this pair gives panel (b), a double-$q$ domain. A fictitious diffraction experiment would observe the same principal magnetic satellite reflections from the pair of single-$q$ domains as from the double-$q$ domain. Thus it is non-trivial to distinguish between these two ordering scenarios, and the same applies when comparing (in three-dimensions) other domain possibilities. Knowledge of the single- or multi-$q$ nature of AFM ordering is important in different areas of condensed matter physics, e.g., in unconventional superconductivity\cite{Yasuyuki_12} and in the understanding of spin-wave dynamics,\cite{Lim_13} as well as in the study of multiferroics.\cite{Ivanov_12}

Traditionally the question as to the single- versus multi-$q$ nature of the AFM domains in high symmetry systems has been addressed by examining, via neutron diffraction, a single-crystal specimen's response to an external perturbation (applied magnetic field or uniaxial stress) that lifts the magnetic degeneracy of symmetry equivalent crystallographic directions.\cite{Rossat_chapter_expt_meths,Forgan_1990,Normile_02} More recently, other scattering approaches that do {\em not} involve an applied perturbation were shown capable, in certain circumstances, of determining the AFM domain nature.\cite{Longfield_02,Stewart_04,Schweizer_08,Stunault_09} In the present study we evidence double-$q$ AFM ordering in a tetragonal crystal system (GdNi$_{\mathrm{2}}$B$_{\mathrm{2}}$C), achieving this result without perturbing the system and through an approach different from those in Refs. \onlinecite{Longfield_02,Stewart_04,Schweizer_08,Stunault_09}.

Our motivation to study GdNi$_{\mathrm{2}}$B$_{\mathrm{2}}$C is threefold: (i) it is justified by the ongoing interest in the rare-earth quaternary borocarbides, which, apart from the well-known superconductivity-magnetism interplay,\cite{Canfield_PhysToday98,Gupta-AdvPhys06} stems from magnetic phenomena in the family that are interesting in their own right;\cite{Alleno_10,ElMassalami_12} (ii) we test a state of the art magnetic structure calculation for rare-earth based antiferromagnets in which the spin (S) is the only contribution to the local magnetic moment;\cite{Jensen_Rotter_08} (iii) [and concomitant with (ii)] we address a paradox\cite{Rotter_06} found in such (4$f$) antiferromagnets. The paradox is as follows. A 4$f$ spin-only system is understood to lack strong single-ion anisotropies derived from crystal fields,\cite{Doerr-AdvPhys05,Rotter_06} and with complexities arising from such anisotropies removed coupled with weak hybridization of the spin polarized electrons with ligand or conduction electrons, the standard model of rare earth magnetism\cite{Jensen-Mackintosh} is expected to provide an accurate description of experiment. However, in several such (4$f$) systems one would expect lattice distortions below their AFM ordering temperatures (N\'{e}el temperatures, T$_N$) and such distortions are {\em not} found experimentally in zero applied magnetic field (H = 0). The term ``magnetoelastic (ME) paradox'' was coined to refer to this inconsistency between experiment and expectation.\cite{Rotter_06}

The expectation of a lattice distortion follows from anticipating the effect of exchange striction (a major component in the standard model\cite{Jensen-Mackintosh} and of importance in fields such as multiferroics\cite{Lee_PRB11}) in each of the experimentally concluded AFM structures.\cite{Rotter_06} Several Gd$^{\text{3+}}$ systems (J = S = 7/2, L = 0), including GdNi$_{\mathrm{2}}$B$_{\mathrm{2}}$C, present the ME paradox, i.e., the experimentally concluded AFM structures are of lower space group symmetry than the lattices, however, the lattices do {\em not} distort to the lower symmetry.\cite{Rotter_06,Rotter_Magelas_para_others_JMMM07} In order to alleviate the ME paradox in GdNi$_{\mathrm{2}}$B$_{\mathrm{2}}$C, Jensen and Rotter undertook model calculations from which they proposed\cite{Jensen_Rotter_08} a double-$q$ magnetic structure with tetragonal symmetry (similar to the lattice) and thus different from the structure concluded from previous scattering studies\cite{Detlefs_96} on GdNi$_{\mathrm{2}}$B$_{\mathrm{2}}$C. Such double-$q$ ordering can be reconciled with the results from the previous scattering studies as being essentially a coherent superposition of the two previously concluded single-$q$ domains.\cite{Detlefs_96,Rotter_06} In the present study we employ resonant elastic x-ray scattering (REXS) to re-examine the magnetic structure of GdNi$_{\mathrm{2}}$B$_{\mathrm{2}}$C, paying particular attention to the possibility of double-$q$ ordering as predicted by Jensen and Rotter.\cite{Jensen_Rotter_08}

\subsection*{Background information and present aims}
GdNi$_{\mathrm{2}}$B$_{\mathrm{2}}$C crystallizes in the tetragonal space group I4/mmm ({\#} 139), with lattice parameters $a$ = $b$ = 3.57 {\AA} and $c$ = 10.37 {\AA}. Magnetization studies\cite{Canfield_95} detect two magnetic phase transitions  upon cooling: long-range AFM order develops at T$_N$ $\approx$ 20 K, and a second, AFM-AFM transition occurs at T$_R$ $\approx$ 14 K. Employing non-resonant and REXS, Detlefs {\em et al.}\cite{Detlefs_96} found the principal magnetic propagation wave vector to be ($q_a$,0,0), or (0,$q_b$,0), with $q_a$ = $q_b$ $\approx$ $\pm$0.55 reciprocal lattice units (rlu). The magnetic ``moment'' direction was reported by these same authors\cite{Detlefs_96} to be in the $a$-$b$ plane and perpendicular to the propagation wave vector (i.e., transversely polarized AFM ordering) down to T$_R$. Below T$_R$ an out of plane ($c$-axis) component associated with the same propagation wave vector,  ($q_a$,0,0) or (0,$q_b$,0), was found to develop, however, the authors\cite{Detlefs_96} did {\em not} determine the phase relationship between the {\em in} and {\em out of} $a$-$b$ {\em plane} components, nor their relative sizes, hence the precise polarization of the low-T AFM order (whether it is, e.g., transverse or elliptical) was {\em not} reported. A neutron powder diffraction by Rotter {\em et al.}\cite{Rotter_06} confirmed the same principal magnetic propagation wave vector, i.e., ($q_a$,0,0) or (0,$q_b$,0), but provided no additional information on the magnetic ordering in GdNi$_{\mathrm{2}}$B$_{\mathrm{2}}$C.

The designation by Detlefs {\em et al.}\cite{Detlefs_96} of ``moment'' direction rather than of (magnetic) ``Fourier component'' direction lay in those authors' assumption of a single-$q$ scenario, in which the ordering wave vector breaks the equivalence of the $a$- and $b$-axes. The symmetry of such an assumed magnetic structure is thus orthorhombic, however, {\em no} signs of any lattice distortion (i.e., any deviation from tetragonal symmetry) were subsequently observed in GdNi$_{\mathrm{2}}$B$_{\mathrm{2}}$C (ME paradox).\cite{Rotter_06} A double-$q$ scenario is predicted by Jensen and Rotter through Landau mean-field theory as well as by numerical mean-field calculations\cite{Jensen_Rotter_08} (see Appendix \ref{Appendix_numerical_calcs}). Jensen and Rotter explain how the double-$q$ scenario leads to a smaller site variation in $|$$<$$\mathbf{J}_i$$>$$|$ ($\propto$ ``ordered moment'') than single-$q$ ordering, implying that GdNi$_{\mathrm{2}}$B$_{\mathrm{2}}$C should stabilize into a double-$q$ structure on similar grounds to those explaining double-$q$ order in cubic compound CeAl$_{\mathrm{2}}$. The numerical calculations reproduce the main features of the magnetic phase diagram of GdNi$_{\mathrm{2}}$B$_{\mathrm{2}}$C previously determined by single-crystal magnetization studies,\cite{Massalami_03} a comparison that adds support to the prediction. Furthermore, the double-$q$ model carries no expectation of a lattice distortion (at H = 0). Hence it offers an explanation for the ME paradox.\cite{Jensen_Rotter_08}

The essential difference between the single- and double-$q$ scenarios is illustrated in Fig. \ref{schematics_1q_2q} (arrows now represent spins of Gd ions) albeit that in this figure the (principal) wave vector is commensurate $q$ = 0.5 rlu (as opposed to $q$ $\approx$ 0.55 rlu) and there is {\em no} out of plane component in these schematics. With regard to evidencing the model in a scattering experiment involving {\em no} external perturbation, an important point is that the Fourier transform of the numerically calculated double-$q$ structure contains ``mixed-order'' Fourier components associated with wave vectors ($nq_a$,$m q_b$,0), with $n$ and $m$ being integers of value 1 or 2, with $n$ $\neq$ $m$. The amplitudes of such components are readily available from the model (see Appendix \ref{Appendix_numerical_calcs}). An aim of the present study is to detect such mixed-order Fourier components via REXS and to thus provide experimental evidence for the double-$q$ model. In the single-$q$ structure, the ($q_a$,0,0) and (0,$q_b$,0) modulations exist in separate domains such that the magnetic structure cannot contain ``mixed-order'' Fourier components. A further aim in the present study is to establish the type of polarization associated with the magnetic ordering. The model calculations\cite{Jensen_Rotter_08} find the AFM ordering below T$_R$ to be elliptically polarized (see Appendix \ref{Appendix_numerical_calcs}).

\section{Experimental Details}
REXS occurs when the incident x-ray photon energy is tuned close to the binding energy of a core level electron, i.e., to an absorption edge.\cite{Hill_McMorrow,Lovesey_Collins} In the hard x-ray range, large resonances are observed from Gd-based magnetic materials at the (Gd) L$_2$ and L$_3$ edges$-$i.e., at the binding energies of Gd 2$p_{1/2}$ and 2$p_{3/2}$ electrons, respectively$-$where the leading order transitions are electric dipole (E1) in nature, viz., the virtual photoelectron probes the unoccupied Gd 5$d$ states.\cite{McMorrow_Handbook} The magnetic origin of such resonant scattering arises when these 5$d$ states carry spin and/or orbital polarization due to intra-ion exchange interaction between the 4$f$ orbitals and 5$d$ band. The resonant part of the Detlefs {\em et al.} study\cite{Detlefs_96} focussed on such E1 REXS and we focus on the same scattering mechanism in the present study.

The single crystal sample of GdNi$_{\mathrm{2}}$B$_{\mathrm{2}}$C studied here was (like the crystal studied in the previous synchrotron studies\cite{Detlefs_96}) grown at the Ames Laboratory using the high-temperature flux technique.\cite{Canfield_R1221_growth} The sample has a platelet form with a large, flat surface, of area 2 $\times$ 2 mm$^2$, perpendicular to the $c$-axis. All REXS measurements have been performed at the {\em XMaS} (BM28) beamline\cite{Brown_XMaS} (ESRF), with the incident x-ray photon energy tuned in the vicinity of the Gd L$_2$ absorption edge. As well as studying in a vertical scattering plane (incident x-ray polarization perpendicular to plane, i.e., $\sigma$ polarized), measurements have also been conducted in a horizontal scattering geometry ($\pi$ polarized incident x-rays); see Figs. \ref{fig_scans}(a) and \ref{Fig_GNBC_Az_T_deps_AsymR_geomSP}(a), where the scattering vector $\mathbf{Q}$ = $\mathbf{k}^{\prime}-\mathbf{k}$, with $\mathbf{k}$  and $\mathbf{k}^{\prime}$ being the incident and exit x-ray wave vectors, respectively. In the horizontal geometry, the dependence of scattering intensity upon rotation of the sample about $\mathbf{Q}$$-$i.e., the azimuthal ($\psi$) dependence$-$has been investigated. A Joule-Thomson cryostat has been used for sample cooling. Polarization analysis of the scattered x-rays has been carried out using a pyrolytic graphite analyzer crystal.

\section{Results and discussion}
\begin{figure}[!hbt]
\includegraphics[scale=.21,angle=0]{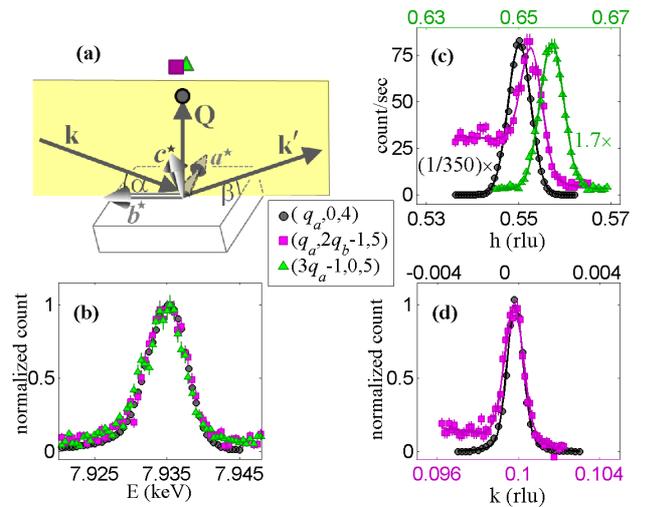}
\caption{(a) Schematic of the vertical scattering geometry used for the comparative study of first- and higher-order satellites: the $\mathbf{b}^\star$ reciprocal lattice axis lies in the scattering plane (yellow shaded region) in all measurements and $\mathbf{c}^\star$ is along the sample surface normal. (b)-(d) Data taken at T = 3 K in $\sigma$$\rightarrow$$\pi^{\prime}$: (b) scans (at fixed $\mathbf{Q}$) as a function of the incident x-ray energy, and (c) {\&} (d) scans parallel to the [100] and [010] crystal axes, respectively. The symbols (black circle, magenta square and green triangle) in (a) indicate the different satellite reflection positions in $\mathbf{Q}$ space and refer to same (central) legend as the plots. The upper abscissa in (c) [(d)] gives the h [k] Miller index corresponding to the scan of the ($\bar{1}$$+$$3q_a$,0,5) [($q_a$,0,4)]. Solid lines are peak fits to Gaussian [(c)] and  pseudo-Voigt [(d)] line shapes. An $\arctan$ function is included to fit the step-like background in the scan of the ($q_a$,$\bar{1}$$+$$2q_b$,5) satellite in (c).}
\label{fig_scans}
\end{figure}

Higher-order satellite reflections are found below T$_N$ at the positions ($q_a$,$\bar{1}$$+$$2q_b$,5) and ($\bar{1}$$+$$3q_a$,0,5), respectively, where $q_a$ = $q_b$ $\approx$ 0.55 rlu. In Fig. \ref{fig_scans} ``energy (E) at-fixed-$\mathbf{Q}$'' and reciprocal space scans of these higher-order reflections, performed in the $\sigma$$\rightarrow$$\pi^{\prime}$ scattering channel at a sample temperature T = 3 K, are compared with similar measurements of the first-order satellite at ($q_a$,0,4). The same resonant character observed for the higher-order satellites as for the first-order satellite$-$panel (b)$-$supports a common magnetic origin of the signals (the common peak position, E$_0$ = 7.9355 keV, is the incident x-ray energy at which the reciprocal space scans have been performed).

The predicted {\em double-q} AFM structure\cite{Jensen_Rotter_08} is composed by a spectrum of Fourier components that includes precisely higher-order components at the wave vectors ($q_a$,2$q_b$,0), with $q_b$ = $q_a$, and (3$q_a$,0,0).  As already mentioned, a Fourier component (hence the observation of a satellite) at a mixed-order wave vector such as ($q_a$,2$q_b$,0) is not expected in the single-$q$ scenario but is critical for the verification of the double-$q$ prediction. No search for higher-order satellites was reported by Detlefs {\em et al.},\cite{Detlefs_96} while in the powder neutron diffraction measurements of Rotter {\em et al.}\cite{Rotter_06} higher-order satellites would not have been visible above the background level (owing to their weakness). Analysis of $^{155}$Gd M\"{o}ssbauer spectra taken on GdNi$_{\mathrm{2}}$B$_{\mathrm{2}}$C showed improved data fitting by the inclusion of a third-order Fourier component.\cite{TomalaPRB98} Since a single-$q$ scenario was assumed in that analysis (as at that time the theory in Ref. \onlinecite{Jensen_Rotter_08} was not available), the effect of including mixed-order Fourier components was {\em not} investigated.

\begin{figure}[!hbt]
\includegraphics[scale=.36,angle=0]{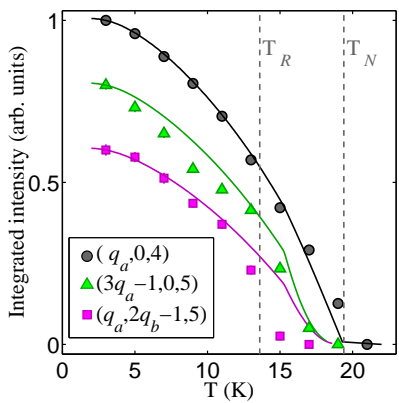}
\caption{Temperature dependence of integrated intensity for the first- and higher-order satellites. The solid lines are simulations based on calculated Fourier components from the double-$q$ model (see Appendices \ref{Appendix_numerical_calcs} and \ref{estimate_rel_intensities}). For clarity, data and simulation have been normalized at T = 3 K to unity  [($q_a$,0,4)], to 0.8 [($\bar{1}$$+$$3q_a$,0,5)], and to 0.6 [($q_a$,$\bar{1}$$+$$2q_b$,5)]. The vertical dashed lines indicate the magnetic transition temperatures from previous studies (see text).} \label{Fig_GNBC_Tdeps_intensities}
\end{figure}

Figure \ref{Fig_GNBC_Tdeps_intensities} shows the temperature dependence of the integrated intensity of each signal (from Fig. \ref{fig_scans}), measured upon sample heating by [100] scans, as well as by both [100] and [010] scans in the mixed-order satellite case. For clarity, each temperature dependence has been normalized to a different intensity value at base temperature. The data are compared with simulations (solid lines) based on the temperature dependence of the corresponding Fourier components of the calculated double-$q$ structure (see Appendices \ref{Appendix_numerical_calcs} and \ref{estimate_rel_intensities}). The ($q_a$,2$q_b$,0) satellite is found to onset at a slightly lower temperature than calculation predicts, however, in general reasonable qualitative agreement between our data and the theoretical simulations is observed, constituting evidence in support of the double-$q$ ordering scenario. We note that the step-like form of the background in the [100] scan through the ($q_a$,$\bar{1}$$+$$2q_b$,5) position$-$Fig. \ref{fig_scans}(c)$-$is found to persist above the onset temperature of this reflection.

The values of the transition temperatures T$_N$ and T$_R$ determined in the Detlefs {\em et al.} study\cite{Detlefs_96} are indicated by the vertical lines in Fig. \ref{Fig_GNBC_Tdeps_intensities}; these values are 19.4 K and 13.6 K, respectively, and magnetization\cite{Canfield_95} and specific-heat\cite{Godart_98} measurements find similar values. As pointed out in Jensen and Rotter's article,\cite{Jensen_Rotter_08} model calculations find a very similar value of T$_N$, however, the calculated value of T$_R$ is around 1.5 K higher than the experimental value.

The relative intensities of the different satellites are indicated in Fig.  \ref{fig_scans}(c), where we plot the true count rate after scaling the first (third) order signal down (up) by a factor of 350 (1.7). In the other plots in this figure, intensities have been  normalized after making a flat background correction to each higher-order satellite scan. Sample rocking scans (not shown) made of the ($q_a$,$\bar{1}$$+$$2q_b$,5) and ($\bar{1}$$+$$3q_a$,0,5) reflections at T = 3 K  for the sample azimuthal orientation indicated in Fig. \ref{fig_scans}(a)$-$i.e., with the $\mathbf{b}^\star$ reciprocal lattice axis lying in the vertical scattering plane for each satellite measurement$-$yield integrated intensities of 0.21 {\%} and 0.12 {\%}, respectively, of the integrated intensity of the sample rocking scan of the ($q_a$,0,4) satellite measured at the same temperature and azimuthal orientation. In the given scattering geometry$-$Fig. \ref{fig_scans}(a)$-$one would expect the (calculated) mixed-order Fourier component to give rise to scattering at ($q_a$,$\bar{1}$$+$$2q_b$,5) that is around two orders of magnitude weaker than that due to the first-order component measured at ($q_a$,0,4), and the scattering due to the third-order component at ($\bar{1}$$+$$3q_a$,0,5) would be weaker still, by a factor close to four (see Appendix \ref{estimate_rel_intensities}). The measured relative integrated intensities of the ($q_a$,$\bar{1}$$+$$2q_b$,5) and ($\bar{1}$$+$$3q_a$,0,5) reflections point to weaker relative scattering strengths compared to theory, by factors of approximately five and three, respectively, which in turn would imply corresponding Fourier components of factors around 2.2 and 1.7, respectively, smaller than calculation. Such discrepancy could be due at least in part to the choice of interaction parameters in Jensen and Rotter's model.\cite{Jensen_Rotter_08} In addition, experimental uncertainty may partially account for the discrepancy; namely, upon comparing intensities of different reflections measured with synchrotron x-rays from the single crystal sample, variations in the sample scattering volume upon changes in the sample orientation may {\em not} be reasonably accounted for by the simple geometric factors$-$viz., $A = 1/(1 + \frac{\sin\alpha}{\sin\beta})$ and $B = \sin\alpha$$-$described in Appendix \ref{estimate_rel_intensities}.

\begin{figure}[!hbt]
\includegraphics[clip=,scale=.21,angle=0]{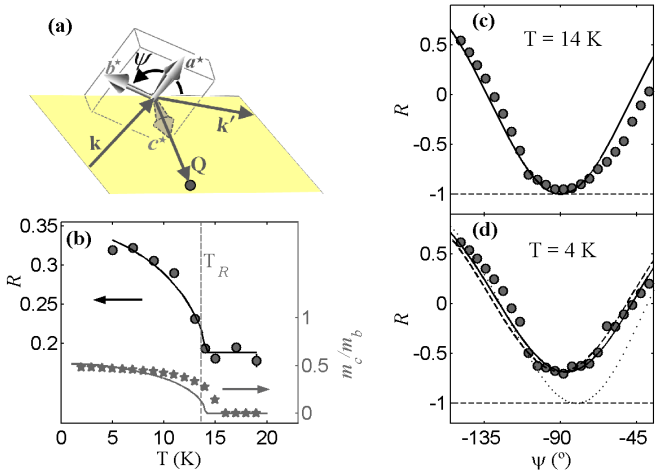}
\caption{(a) Horizontal scattering geometry used to study the azimuthal ($\psi$) dependence of the ($q_a$,0,6) satellite scattering. (b) Temperature dependence of $R$ [Eq. (\ref{asymm_rat})] measured for the sample orientation ($\psi$ setting) depicted in panel (a). The lower (gray) solid line in (b) gives the $m_c/m_b$ values obtained from the fit to $R$(T) [the upper black solid line], and the stars are the $m_c/m_b$ values from model calculations (both curves refer to the right axis). (c) {\&} (d) $\psi$-dependence of $R$ measured above and below  T$_R$, respectively. The different ``fits'' in (d) are described in the text.} \label{Fig_GNBC_Az_T_deps_AsymR_geomSP}
\end{figure}

We move now to the determination of the polarization of the AFM ordering. The model calculations\cite{Jensen_Rotter_08} find the ordering to be elliptically polarized below T$_R$, which corresponds to a phase relationship of $e^{\pm i \pi/2}$ between projections onto the $b$ [$a$] and $c$ axes of the first-order Fourier component with wave vector ($q_a$,0,0) [(0,$q_b$,0)]; see Appendix \ref{Appendix_numerical_calcs}. We show in Fig. \ref{Fig_GNBC_Az_T_deps_AsymR_geomSP} the results from our study, in horizontal scattering, of the first-order satellite positioned at ($q_a$,0,6). Measuring sample rocking curves of the satellite (at resonance) in both the $\pi$$\rightarrow$$\sigma^{\prime}$ and $\pi$$\rightarrow$$\pi^{\prime}$ scattering channels, the asymmetry ratio
\begin{equation}
R = \frac{I^{(\pi\sigma^{\prime})} - I^{(\pi\pi^{\prime})}}{I^{(\pi\sigma^{\prime})} + I^{(\pi\pi^{\prime})}}
\label{asymm_rat}
\end{equation}
(where $I$ denotes integrated intensity) has been determined as a function of temperature at fixed azimuthal angle ($\psi$ = $-140^{\circ}$), panel (b), and as a function of $\psi$ at fixed temperatures; T = 4 K ($<$ T$_R$), panel (c), and T = 14 K ($\gtrsim$ T$_R$), panel (d). The $\psi$ angle is defined with respect to the $\mathbf{b}^\star$ axis and is zero when this axis lies in the scattering plane, on the exit beam side. The $\psi$ angle shown in Fig. \ref{Fig_GNBC_Az_T_deps_AsymR_geomSP}(a) is negative. A positive change in $\psi$ rotates the sample clockwise about $\mathbf{Q}$.

From the scattering amplitudes for $\pi$$\rightarrow$$\sigma^{\prime}$ and $\pi$$\rightarrow$$\pi^{\prime}$  $E1$  scattering,\cite{Hill_McMorrow,Lovesey_Collins} we find (by writing the amplitudes in terms of Fourier components, as described in Appendix \ref{estimate_rel_intensities} for the case of $\sigma$$\rightarrow$$\pi^{\prime}$ $E1$ scattering) that the measured asymmetry ratio should conform to the following function
\begin{equation}
R_{\text{fit}} = \frac{|\mathbf{m}_{\mathbf{q}_{\text{first}}}.\hat{\mathbf{k}}|^2 - |\mathbf{m}_{\mathbf{q}_{\text{first}}}.(\hat{\mathbf{k}^{\prime}} \times \hat{\mathbf{k}})|^2}{|\mathbf{m}_{\mathbf{q}_{\text{first}}}.\hat{\mathbf{k}}|^2 + |\mathbf{m}_{\mathbf{q}_{\text{first}}}.(\hat{\mathbf{k}^{\prime}} \times \hat{\mathbf{k}})|^2}
\label{asymm_rat_fit}
\end{equation}
where the two scalar products are, of course, functions of $\psi$. The fitting curves in Fig. \ref{Fig_GNBC_Az_T_deps_AsymR_geomSP} are based on this equation. The function is insensitive to the magnitude of the first-order Fourier component ($\mathbf{m}_{\mathbf{q}_{\text{first}}}$), since this magnitude cancels between the numerator and denominator, hence the fits determine the unit vector $\hat{\mathbf{m}}_{\mathbf{q}_{\text{first}}}$ = [0,$m^{\prime}_{\mathbf{q}_{\text{first}},b}$,$m^{\prime}_{\mathbf{q}_{\text{first}},c}$], with $m^{\prime}_{\mathbf{q}_{\text{first}},b}$ = $1/(1+(m_c/m_b)^2)^{1/2}$ and $m^{\prime}_{\mathbf{q}_{\text{first}},c}$ = $(m_c/m_b)e^{i\phi}m^{\prime}_{\mathbf{q}_{\text{first}},b}$. There are two adjustable parameters controlling each fitted value of $R$: (i) the ratio $m_c/m_b$ and (ii) the phase angle $\phi$.

The fit in Fig. \ref{Fig_GNBC_Az_T_deps_AsymR_geomSP}(c) involves no adjustable parameters: $m_c/m_b$ (and, hence, $m^{\prime}_{\mathbf{q}_\text{1},c}$) is fixed to zero. The achievement of a fit to the data in panel (d) is sensitive to the value of $\phi$. The solid, dashed and dotted line curves correspond to fits with $\phi$ floated, fixed at $\pi/2$ and fixed at 0 (or $\pi$), respectively. The fitted value is $\phi$ = 1.42(5) rad, in good agreement with the theoretical value of $\pi/2$ (see Appendix \ref{Appendix_numerical_calcs}). We should note that the analysis of $^{155}$Gd M\"{o}ssbauer spectra taken on GdNi$_{\mathrm{2}}$B$_{\mathrm{2}}$C (Ref. \cite{TomalaPRB98}) suggested such elliptical polarization. Here we find definitive evidence for this type of AFM polarization from our scattering experiment. The ratio $m_c/m_b$ was floated in all fits in Fig. \ref{Fig_GNBC_Az_T_deps_AsymR_geomSP}(c), producing a value of 0.53(2) in the ``$\phi$ floated'' fit, in good agreement with the value (0.48) from model calculations for the same temperature (T = 4 K). The fit to the temperature dependence of $R$$-$upper solid line in panel (b)$-$is with $m_c/m_b$ constrained to follow a ``J = 7/2 mean-field'' temperature dependence and $\phi$ fixed to 1.42 rad. We use this functional form to extract a smooth curve describing the experimental variation in $m_c/m_b$ with temperature (gray line plotted below the data and referring to the y-axis on the right), which may be directly compared with the ratio calculated from the double-$q$ model (stars)$-$see Appendix \ref{Appendix_numerical_calcs}. The agreement with theory is reasonable (the discrepancy in the value of T$_R$ between calculation and experiment has already been mentioned above).

\section{Conclusions}
Following the recent prediction of double-$q$ magnetic ordering to alleviate the ME paradox,\cite{Jensen_Rotter_08} we have re-examined the magnetic structure of GdNi$_{\mathrm{2}}$B$_{\mathrm{2}}$C using REXS. The observation of a mixed-order magnetic satellite reflection clearly confirms the hypothesis of a double-$q$ magnetic structure, without the need to apply an external symmetry-breaking perturbation. Our study thus constitutes an example of a ``theory-guided'' approach to the establishment of double-$q$ AFM order in a high symmetry crystal material, complementing other scattering approaches that have evidenced multi-$q$ order {\em without} employing symmetry-breaking perturbations.\cite{Longfield_02,Stewart_04,Schweizer_08,Stunault_09}

The signal strengths and temperature dependencies of the mixed-order as well as of a third-order satellite are in qualitative agreement with theory.\cite{Jensen_Rotter_08} However, the precise intensities of these higher-order satellites with respect to a first-order reflection suggest attempting future refinement of the values of the interaction parameters adopted in the model, in order to investigate whether improved quantitative agreement with experiment may be achieved.

By examining the sensitivity of first-order satellite scattering to sample rotation about the scattering vector, we evidence the theoretically expected elliptical polarization of the magnetic ordering at low temperature, i.e., we find the phase factor $e^{i\phi}$ linking the projections along the $b$ and $c$ axes of the first-order Fourier component at ($q_a$,0,0) to correspond to $\phi$ $\approx$ $\pi/2$. We find the variation with temperature of the ratio of these projections, i.e., $m_c/m_b$, to be in fairly good agreement with the corresponding temperature dependence from numerical calculations based on the double-$q$ model.\cite{Jensen_Rotter_08}

In future scattering studies it would be interesting to determine the magnetic field dependence of the mixed-order satellite to contrast with magnetization studies\cite{Massalami_03} on GdNi$_{\mathrm{2}}$B$_{\mathrm{2}}$C as well as with calculations\cite{Jensen_Rotter_08} for H $\neq$ 0. The calculations\cite{Jensen_Rotter_08} and present REXS results encourage similar (combined) studies to help elucidate the ME paradox in other Gd-based compounds.\cite{Rotter_06}

\begin{acknowledgments}
The EPSRC-funded {\em XMaS} beam line at the ESRF is directed by M.J. Cooper, C.A. Lucas, and T.P.A. Hase. We are grateful to O. Bikondoa, L. Bouchenoire, S. Brown and P. Thompson for their invaluable assistance and to S. Beaufoy and J. Kervin for additional {\em XMaS} support. P.C.C.'s work was supported by the U.S. Department of Energy, Office of Basic Energy Science, Division of Materials Sciences and Engineering. P.C.C.'s synthesis and basic characterization was performed at the Ames Laboratory. Ames Laboratory is operated for the U.S. Department of Energy by Iowa State University under Contract No. DE-AC02-07CH11358. JAB acknowledges financial support from the Spanish  MINECO and a European Regional Development Fund Grant (No. MAT2011-27573-C04-02). MR gratefully acknowledges useful discussions with Maurits Haverkort.
\end{acknowledgments}

\appendix

\section{Numerical mean field calculations: magnetic Fourier components\label{Appendix_numerical_calcs}}
\begin{figure}[!hbt]
\includegraphics[clip=,scale=.2,angle=0]{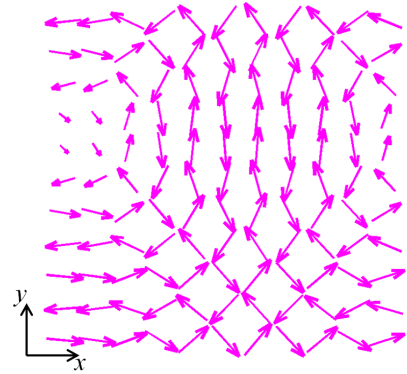}
\caption{Calculated magnetic structure of GdNi$_{\mathrm{2}}$B$_{\mathrm{2}}$C at T = 3 K. The arrows represent projections onto the $xy$ (or $ab$) plane of the spins of an $xy$ plane of Gd ions. } \label{mag_struct_3K}
\end{figure}

Self-consistent mean-field calculations have been made as a function of temperature using the MCPHASE program,\cite{Rotter_McPhase_JMMM04} in accordance with the information given in the Jensen and Rotter paper.\cite{Jensen_Rotter_08} The resulting double-$q$ magnetic structure at T = 3 K (the temperature corresponding to the measurements in Fig. \ref{fig_scans}) is illustrated in Fig. \ref{mag_struct_3K}. The calculated ordered (spin) moment at each Gd site ($\mathbf{r}_n$) has a magnitude that is independent of position, i.e.,  $\left\lvert\mathbf{m}(\mathbf{r}_n)\right\rvert$ = 7 $\mu_B$ for all Gd ion positions, where $\mathbf{m}(\mathbf{r}_n)$ denotes magnetic moment as a function of position. For clarity, in Fig. \ref{mag_struct_3K} we show only  projections of spins onto the $ab$ plane. Where the projection is small, the spin component along $c$ (not shown) is large, thus providing the constant $\left\lvert\mathbf{m}(\mathbf{r}_n)\right\rvert$.

Fourier components, $\mathbf{m}_{\mathbf{q}}$, of such calculated structures have been computed using the MCPHASE program. We adopt the following definition of $\mathbf{m}_{\mathbf{q}}$
\begin{equation}
\mathbf{m}(\mathbf{r}_n) = \sum_{\mathbf{q}} \mathbf{m}_{\mathbf{q}}e^{-i\mathbf{q}.\mathbf{r}_n}
\label{Fourier_comp}
\end{equation}
For T = 3 K, the calculated Fourier component at the wave vector ($q_a$,0,0) has projections onto the $a$, $b$ and $c$ axes of zero, ($-2.70 - 1.77i$) $\mu_B$ and ($1.15 - 1.75i$) $\mu_B$, respectively. The phase angle between these $b$ and $c$ projections is $\pi/2$, and the same angle is found for other temperatures below T$_R$. Such a phase angle implies elliptically polarized AFM ordering at low temperature.

For T $<$ T$_R$ the calculated double-$q$ structure is asymmetric. This is evidenced by, for example, the result that in this temperature range the Fourier component at the wave vector (0,$q_b$,0) is of a different magnitude from that at ($q_a$,0,0). At T = 3 K, the former component  has projections onto the $a$, $b$ and $c$ axes of ($-0.415 + 2.90i$) $\mu_B$, zero and ($0.0154 + 0.0022i$) $\mu_B$, respectively, where again the phase angle between the non-zero projections (this time along $a$ and $c$) is $\pi/2$. Given the tetragonal lattice, this asymmetry implies the formation of two types of orientational double-$q$ (AFM) domains within a single crystal specimen. Namely, for example, the above values of  Fourier components at the wave vectors ($q_a$,0,0) and (0,$q_b$,0) will be interchanged from grain to grain, and the same applies to  components at other symmetry equivalent wave vectors.

Since in a diffraction experiment, the scattering probe will illuminate a {\em large} number of such domains, the magnetic scattering signal will effectively average, in a geometrically fashion, over different domains. In Table \ref{table1} we give geometrical averages of the calculated Fourier components relevant to the measurements in Fig. \ref{fig_scans}.

\begin{table}[!hbt]
\begin{center}
\caption{Projections onto the three principal crystallographic axes of calculated Fourier components at the given propagation wave vectors for T = 3 K. We give absolute values ($\left\lvert\mathbf{m}_{\mathbf{q}}\right\rvert$) only, which are geometrical averages over the two predicted double-$q$ orientational domains (see text).}
\begin{tabular}{ccccccc}
\hline \hline
 & \vline & \multicolumn{5} {c}  {Fourier component projections ($\mu_B$)}    \\
Wave vector & \vline & \hspace{2.5mm} {$a$-axis} \hspace{2.5mm} & \vline & \hspace{2.5mm} {$b$-axis} \hspace{2.5mm} & \vline & \hspace{2.5mm} {$c$-axis} \hspace{2.5mm}   \\
\hline
$\mathbf{q}_{\text{first}}$ = ($q_a$,0,0) & \vline & 0  & \vline & 3.08  & \vline&  1.48   \\
$\mathbf{q}_{\text{mixed}}$ = ($q_a$,$2q_b$,0)& \vline & 0.0046   & \vline& 0.340   & \vline& 0.223   \\
$\mathbf{q}_{\text{third}}$ = ($3q_a$,0,0) & \vline & 0  & \vline &  0.177  & \vline & 0.089    \\
\hline \hline
\end{tabular}
\label{table1}
\end{center}
\end{table}

\section{Relative satellite intensities estimated from calculated Fourier components \label{estimate_rel_intensities}}
The $E1$ scattering amplitude relevant to the present study is normally expressed in terms of $\hat{\mathbf{z}}_n$, the unit vector pointing along the direction of the $n$th magnetic moment, i.e., $\hat{\mathbf{z}}_n$ = $\mathbf{m}(\mathbf{r}_n)/\left\lvert\mathbf{m}(\mathbf{r}_n)\right\rvert$. Here we express it in terms of a sum over magnetic Fourier components ($\mathbf{m}_{\mathbf{q}}$), following a similar approach to that taken to magnetic structure factors in the analysis of neutron diffraction data from multi-$q$ systems.\cite{Rossat_chapter_expt_meths}

The $\sigma$$\rightarrow$$\pi^{\prime}$ $E1$ resonant scattering amplitude of the (in our case) Gd ion at the $n$th crystallographic site, located at the position $\mathbf{r}_n$, is given by $f^{(\sigma\pi^{\prime})}_{n}$ = $iF^{(1)}\hat{\mathbf{k}^{\prime}}.\hat{\mathbf{z}}_n$, where $F^{(1)}$ is a difference between matrix element-based terms ($F_{LM}$)$-$see Refs. \onlinecite{Hill_McMorrow} and \onlinecite{Lovesey_Collins}. From Eq. (\ref{Fourier_comp}) we may write
\begin{equation}
f^{(\sigma\pi^{\prime})}_{n} = iF^{(1)}\frac{\sum_{\mathbf{q}} (\mathbf{m}_{\mathbf{q}}.\hat{\mathbf{k}^{\prime}})e^{-i\mathbf{q}.\mathbf{r}_n}}{\left\lvert\mathbf{m}(\mathbf{r}_n)\right\rvert}
\label{Pol_sp_Fourier}
\end{equation}
Substituting this into a structure factor, i.e., $\sum_{n} f^{(\sigma\pi^{\prime})}_{n}e^{i\mathbf{Q}.\mathbf{r}_n}$ with the sum running over an entire AFM domain volume, and considering the case of a site-independent value of $\left\lvert\mathbf{m}(\mathbf{r}_n)\right\rvert = M$ (as is found for the calculated structure mentioned above), we may write
\begin{equation}
\sum_{n} f^{(\sigma\pi^{\prime})}_{n}e^{i\mathbf{Q}.\mathbf{r}_n} = \frac{iF^{(1)}}{M}\sum_{\mathbf{q}}\sum_{n}(\mathbf{m}_{\mathbf{q}}.\hat{\mathbf{k}^{\prime}})e^{i(\mathbf{Q}-\mathbf{q}).\mathbf{r}_n}
\label{Pol_sp_Fourier2}
\end{equation}
Each summation over $n$ vanishes for every wave vector $\mathbf{q}$ except that corresponding to the given Bragg condition, $\mathbf{Q}$ = $\boldsymbol{\tau}_{\ast} + \mathbf{q}_{\ast}$, where $\boldsymbol{\tau}$ denotes a reciprocal lattice vector and the subscript $\ast$ indicates a specific vector. Thus
 \begin{equation}
 \sum_{\mathbf{q}}\sum_{n} \mathbf{m}_{\mathbf{q}}.\hat{\mathbf{k}^{\prime}} e^{i(\mathbf{Q}-\mathbf{q}).\mathbf{r}_n} = \sum_{\mathbf{q}}\mathbf{m}_{\mathbf{q}}.\hat{\mathbf{k}^{\prime}}\sum_{n} \delta_{\mathbf{Q}-\mathbf{q},\boldsymbol{\tau}_{\ast}}
\label{SF_Kronecker_delta}
\end{equation}
where the Kronecker delta $\delta_{\mathbf{Q}-\mathbf{q},\boldsymbol{\tau}_{\ast}}$ $\equiv$ $\delta_{\mathbf{q},\mathbf{q}_{\ast}}$, and hence
\begin{equation}
\sum_{n} f^{(\sigma\pi^{\prime})}_{n}e^{i\mathbf{Q}.\mathbf{r}_n} \propto \frac{iF^{(1)}\mathbf{m}_{\mathbf{q}_{\ast}}.\hat{\mathbf{k}^{\prime}}}{M}
\label{SF_sp_Fourier}
\end{equation}
where $\mathbf{m}_{\mathbf{q}_{\ast}}$ denotes the {\em specific} Fourier component being {\em sampled} (``filtered out'') by the REXS process at the given photon momentum transfer vector $\mathbf{Q}$. Since the terms $iF^{(1)}$ and $M$ are the same for any given satellite measurement, be it first- or higher-order, the intensity ($\propto |${\em structure factor}$|^2$) simulations in Fig. \ref{Fig_GNBC_Tdeps_intensities} have been evaluated simply as $|\mathbf{m}_{\mathbf{q}_{\ast}}.\hat{\mathbf{k}^{\prime}}|^2$, using geometrically averaged Fourier components calculated as a function of temperature (see above).

With regard to the integrated intensities of rocking curves of the different satellites, three experimental factors affecting the integrated intensity have been taken into account: (i) the Lorentz factor, $L = 1/\sin2\theta$, where $2\theta$ is the scattering angle; (ii) the factor $A = 1/(1 + \frac{\sin\alpha}{\sin\beta})$, related to the x-ray attenuation by the sample; and (iii) $B = \sin\alpha$, the incident beam fraction intercepted by the sample. The angles $\alpha$ and $\beta$ are defined in Fig. \ref{fig_scans}(a), and their values during the first-, mixed- and third-order measurements are [($\alpha$,$\beta$) $\approx$] (18.9$^\circ$,18.9$^\circ$), (20.2$^\circ$,26.5$^\circ$) and (23.7$^\circ$,23.7$^\circ$), respectively. We have evaluated the ratios $(LAB)_{\text{first}}/(LAB)_{\text{mixed}}$ and $(LAB)_{\text{first}}/(LAB)_{\text{third}}$$-$where the subscripts refer to the different (satellite) diffraction conditions, respectively$-$and find them both to be very close to (within 3 {\%} of) unity. Thus, we may compare the relative integrated intensities $I_{\text{mixed}}/I_{\text{first}}$ and $I_{\text{third}}/I_{\text{first}}$ directly to modulus squared values of ratios of calculated structure factors. In calculating these ratios, common factors$-$i.e., $iF^{(1)}$ and $M$ in Eq. (\ref{SF_sp_Fourier})$-$in the numerator and denominator  cancel out. Therefore, the relevant structure factor ratios reduce to $\left\lvert\frac{\mathbf{m}_{\mathbf{q}_{\text{mixed}}}.\hat{\mathbf{k}^{\prime}}}{\mathbf{m}_{\mathbf{q}_{\text{first}}}.\hat{\mathbf{k}^{\prime}}}\right\rvert^2$ and $\left\lvert\frac{\mathbf{m}_{\mathbf{q}_{\text{third}}}.\hat{\mathbf{k}^{\prime}}}{\mathbf{m}_{\mathbf{q}_{\text{first}}}.\hat{\mathbf{k}^{\prime}}}\right\rvert^2$. These ratios have been evaluated for T = 3 K using geometrical averages of the calculated Fourier components (see Table \ref{table1}) and taking into account the sample azimuthal orientation indicated in Fig. \ref{fig_scans}(a). The resulting ratio values are 0.011 and 0.0032, respectively, which are factors of approximately five and three times larger, respectively, than the corresponding experimental relative intensities (0.0021 and 0.0012).


\end{document}